\documentclass{elsart}

\usepackage{amssymb}

\usepackage{graphicx}

\begin{document}

\begin{frontmatter}
\title{Quasi-geostrophic kinematic dynamos at low magnetic Prandtl number}

\author{Nathanael Schaeffer and Philippe Cardin}

\address{LGIT, Observatoire de Grenoble \\
Universit\'e Joseph Fourier, and CNRS\\
Grenoble, France.}

\begin{abstract}
Rapidly rotating spherical kinematic dynamos are computed using the combination of a quasi geostrophic (QG) model for the velocity field and a classical spectral 3D code for the magnetic field. On one hand, the QG flow is computed in the equatorial plane of a sphere and corresponds to Rossby wave instabilities of a geostrophic internal shear layer produced by differential rotation. On the other hand, the induction equation is computed in the full sphere after 
a continuation of the QG flow along the rotation axis.
Differential rotation and Rossby-wave propagation are the key ingredients of the dynamo process which can be interpreted in terms of $\alpha\Omega$ dynamo. Taking into account the quasi geostrophy of the velocity field to increase its time and space resolution enables us to exhibit numerical dynamos with very low Ekman (rapidly rotating) and Prandtl numbers (liquid metals) which are asymptotically relevant to model planetary core dynamos.
\end{abstract}


\end{frontmatter}

\section{Introduction}

The magnetic field of the Earth is produced by a dynamo effect in the metallic liquid core of our rotating planet. Many efforts have been made successfully in the last decade to describe the mechanism of a self induced magnetic field either with experimental models \cite{gai01,sti01,car04} or numerical simulations \cite{gla95,kag97,kua97,chr99}. Both approaches have limitations. No experiment has been done in rotation while rotation is seen as a key ingredient by geophysicists to explain the geometry and the amplitude of the geomagnetic field \cite{gub87}. All numerical models \cite{dor00,kon02} have introduced the Coriolis force in solving the Navier-Stokes equation and the 
quasi geostrophy (two dimensionality imposed by the Taylor Proudman theorem \cite{gre68}) of the flow participates in the generation of the magnetic field. Thermal convective vortices aligned with the rotation axis are associated to surface patches of magnetic field \cite{chr99} and spatio-temporal behaviors of magnetic and vorticity field are similar. This effect is a direct consequence of the prescribed magnetic Prandtl number ($P_m = \nu/\eta$, where $\nu$ is the kinematic viscosity and $\eta$ the magnetic diffusivity) in the simulations. The current computer capacities limit the computation to magnetic Prandtl number of the order of unity \cite{dor00} while liquid metals exhibit magnetic Prandtl number lower than $10^{-5}$, 
even in the planetary core conditions \cite{poi94}.

In this paper, we propose an approach that aims at computing very low magnetic Prandtl number dynamos taking advantage of the quasi-geostrophic behavior of the velocity field.
For very low Ekman number ($ E = \nu/\Omega R^2$, where $\Omega$ is the rotation rate of the spherical container, and $R$ its radius), a quasi-geostrophic (QG) approach models correctly the flow in a rapidly rotating sphere \cite{bus70,car94}.  It consists of the integration of the flow equations along the rotation axis. Even if the numerical resolution is done with a stream function in the equatorial plane (2D), the top and bottom boundary conditions are taken into account through slope ($\beta$) and Ekman pumping effects. In the context of the study of thermal convection in rapidly rotating spherical shells, Aubert et al. \cite{aub03} have compared successfully their QG results with 3D calculations \cite{dor04} and experimental measurement \cite{aub01}.
Low value of $P_m$ may imply a separation in term of scales and frequencies, between the velocity and magnetic fields in a metallic dynamo. This idea has already been applied to kinematic dynamo computations at low $P_m$ \cite{pon04}.

In this work, we compute the QG flow in the equatorial plane with a fine spatio-temporal resolution and the velocity is extrapolated to a coarse 3D spherical grid where the induction equation is solved.

In order to demonstrate the validity of this approach, we have decided to apply it to a simple case. Instead of a thermal convective flow for which heat transport has to be modeled, we model the instabilities of an internal geostrophic shear layer. This layer, known as the Stewartson layer, is produced by a differentially rotating inner core in a rotating sphere and consists of two nested viscous shear layers \cite{ste66,dor98}. For a critical Ro number ($ Ro = \Delta \Omega / \Omega$, where $\Delta \Omega$ is the differential rate of rotation of the inner core), the Stewartson layer becomes unstable \cite{hol03} and generates Rossby waves \cite{sch04}.

As we will show in this paper, such kind of flow can generate and sustain a magnetic field.
The QG-model allows us to compute dynamos at very low Ekman (down to $10^{-8}$) and Prandtl numbers (as low as $3\,10^{-3}$).

\section{The equations}

\begin{figure}
	\hspace{25pt}\includegraphics[width=6cm]{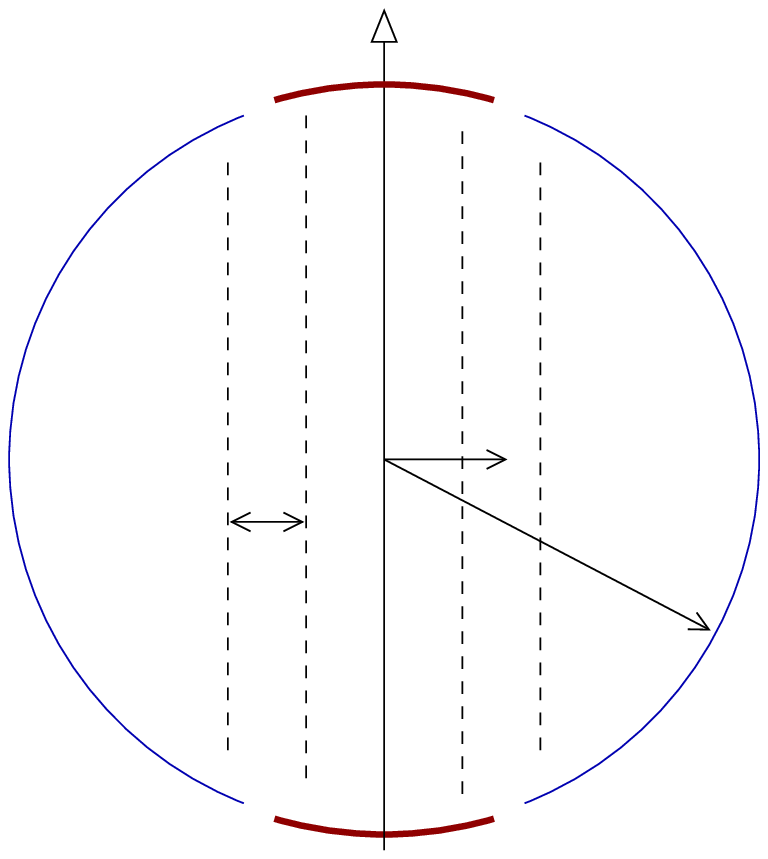} \\
	\begin{picture}(0,0)(-110,-180)
		\put(-100,-10){shear layer}
		\put(-50,-15){\vector(1,-1){25}}
		\put(15,-70){$R_s$}
		\put(50,-100){$R$}
		\put(-25,-105){$E^{1/4}$}
		\put(-65,-70){$\Omega$}
		\put(60,-70){$\Omega$}
		\put(-20,-25){$\Omega+\Delta\Omega$}
		\put(7,5){$z$}
	\end{picture}
\vspace{-0.5cm}
\caption{Sketch of the split sphere geometry. The differential rotation produces an axisymmetric Stewartson $E^{1/4}$ shear layer which is cylindrical and aligned with the rotation axis $z$.}
	\label{fig:geom}
\end{figure}

\subsection{Hydrodynamics}

Let us consider a sphere of radius $R$ filled with an incompressible liquid metal of viscosity $\nu$ and magnetic diffusivity $\eta$. The sphere is rotating at $\Omega$ along the z-axis of a cylindrical reference frame (${\bf e_s}, {\bf e_{\phi}}, {\bf e_z}$). The sphere is split at the colatitude $\pm \sin^{-1}(R_s/R)$ ($R_s/R$ is set to 0.35). The two polar caps are differentially rotating  at $\Delta \Omega$
as shown in figure \ref{fig:geom}.
 $\Omega^{-1}$ is chosen to scale the time, $R$ the length, $(\mu_0 \rho)^{1/2} R \Omega$ the magnetic field. For low Ekman and Rossby numbers, the flow is quasi geostrophic \cite{gre68}. Taking the curl of the Navier-Stokes equation and averaging along the rotation axis z (noted by an overbar), we get the QG equation for the z-component of the vorticity $\omega = {\bf e_z} \cdot \nabla\times{\bf u}$, provided that $u_s$ and $u_{\phi}$ are independent of $z$ \cite{sch04}.
\begin{equation}
\label{eq:Vort2D}
        \frac{\partial \omega}{\partial t} 
        + u_s \frac{\partial \omega}{\partial s} 
        + \frac{u_{\phi}}{s} \frac{\partial \omega}{\partial \phi}
        - (2+\omega) \overline{\frac{d u_z}{d z}} 
        = \overline{\nabla\times\left({\bf j} \times {\bf B}\right)\cdot{\bf e_z}} 
          +E \Delta \omega
\end{equation}
The Coriolis term needs the evaluation of $\overline{\frac{d u_z}{d z}}$. 
We deduce that $u_z$ is a linear function of $z$ from the averaged mass conservation equation. Consequently, its vertical derivative may be deduced from the non penetration boundary condition ($\beta$ effect) and the viscous boundary condition (the Ekman pumping effect) \cite{sch04}. It gives:

\begin{equation}
\label{eq:duzPump}
  \overline{\frac{d u_z}{dz}} = E^{1/2} P(u_s,u_{\phi},s) + \beta(s) u_s
\end{equation}
where $\beta(s) \equiv \frac{1}{L} \left.\frac{d L}{ds}\right|_{z=L}$ and $L(s) = \sqrt{1 -s^2}$ is the half height of a column of fluid and 
\begin{equation}
P(u_s,u_{\phi},s) = \frac{1}{2(1-s^2)^{3/4}} \left[ - \omega + \frac{s}{1-s^2} \left( \frac{\partial u_s}{\partial \phi}
    -\frac{1}{2} u_{\phi} \right) - \frac {5s}{2(1-s^2)^{3/2}} u_s \right]
\end{equation}
is the pumping boundary condition in a rigid sphere deduced from Greenspan's formula \cite{sch04}.

The axisymmetric flow is computed directly from the velocity equation.
\begin{equation}
\label{eq:NLbase} 
    \frac{\partial \left\langle{u}_{\phi}\right\rangle}{\partial t} 
    + \left\langle  u_s \frac {\partial u_{\phi}}{\partial s} \right\rangle         
    + \frac {\left\langle  u_{\phi}u_s\right\rangle}{s} 
 	  + 2\left\langle {u}_s \right\rangle 
 	  = \left\langle {\overline{({\bf j\times B})\cdot{\bf e_{\phi}}}} \right\rangle
 	  + E \left( \Delta \left\langle {u}_{\phi}\right\rangle 
 	            - \frac{\left\langle {u}_{\phi}\right\rangle}{s^2} \right)
\end{equation}
where $\left\langle \; \right\rangle$ stands for the $\phi$-average operator. Rigid boundary conditions are assumed for the velocity at $s=1$. For $s<R_s/R$, the top and bottom azimuthal velocity are imposed as $u_{\phi} = s Ro$.
The velocity field is computed using a generalised stream function in the equatorial plane as in \cite{sch04} which guarantees the 3D mass conservation. The stream function is expanded in Fourier components along the $\phi$ component. 
It may be interesting to introduce the Reynolds number $Re = Ro E^{-1}$ directly related to the two controlling dimensionless numbers $E$ and $Ro$.

In this paper, as a first step, we will only consider kinematic dynamos and the magnetic terms in (\ref{eq:Vort2D}, \ref{eq:NLbase}) will be neglected.

\subsection{Induction equation}

The velocity field computed with equations (\ref{eq:Vort2D}, \ref{eq:NLbase}) in the equatorial plane is extrapolated to a spherical grid (on Gauss collocation points) in the physical space. This is a straightforward process because $u_s$ and $u_{\phi}$ are independent of $z$ and $u_z$ is a linear function of $z$. Then, the velocity field is changed into spherical coordinates (${\bf e_r},{\bf e_{\theta}}, {\bf e_{\phi}}$) to compute the non linear induction term. The dimensionless equation of evolution of the magnetic field is :
\begin{equation}
\label{eq:ind} 
    \frac{\partial B}{\partial t} = \nabla\times\left(u\times B \right) 
 	  + P_m^{-1} E \Delta B 
\end{equation}
Changes of magnetic Prandtl number $P_m^{-1}$ would change directly the magnetic Reynolds number $R_m = Re P_m = Ro E^{-1} P_m$ which is more commonly used in dynamo modeling.
This equation is solved using spherical harmonics where the magnetic boundary conditions are easy to write \cite{gub87}. The induction part of the code has been checked using kinematic dynamo results \cite{dud89} and the dynamo benchmark \cite{chr01}.

\subsection{Numerical implementation}

A finite difference scheme is used on an irregular radial grid (denser in the Stewartson and Ekman layers). A semi implicit Crank-Nicholson scheme is used for linear terms in time whereas an Adams-Bashforth procedure is implemented for non linear terms. For low $P_m$, cylindrical and spherical radial grid steps may differ by a factor $20$. Similarly, time steps for the induction equation may be much longer than the velocity time steps (as much as $20$ times). For a run at $E=10^{-8}$, the stream function is computed on a cylindrical mesh made of 600 radial points and expanded in Fourier series up to degree $m=170$ while the magnetic field is expanded in spherical harmonics ($L_{max} = 79, M_{max} =32$) with an irregular radial grid of 150 points for $P_m = 10^{-2.5}$.
By increasing $M_{max}$ and reducing the time step factor, we checked that such truncatures do not influence the onset of dynamo action.

\section{Hydrodynamics}

\begin{figure}
\begin{center}
	\includegraphics[width=5.8cm]{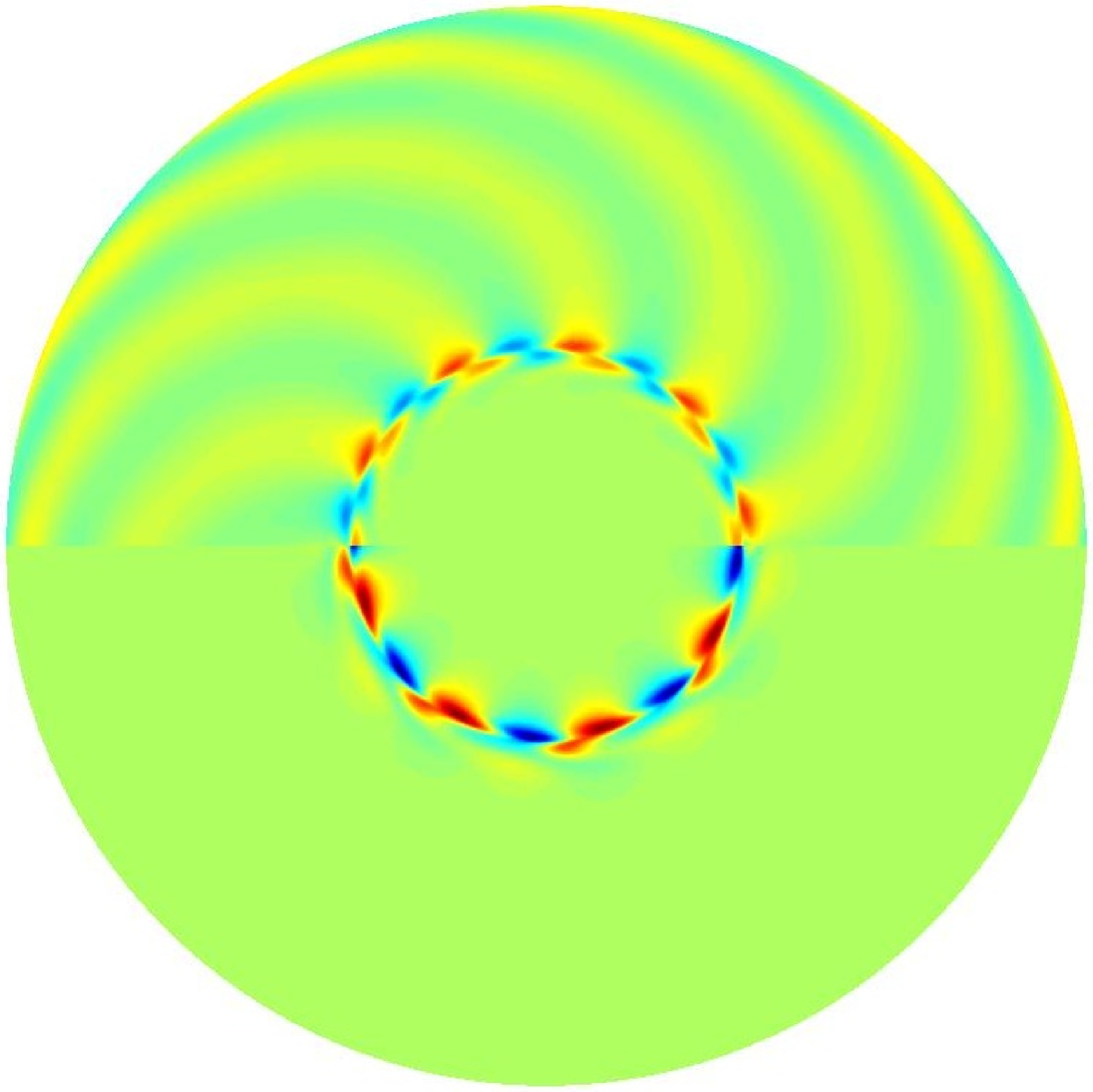}
	\includegraphics[width=6.8cm]{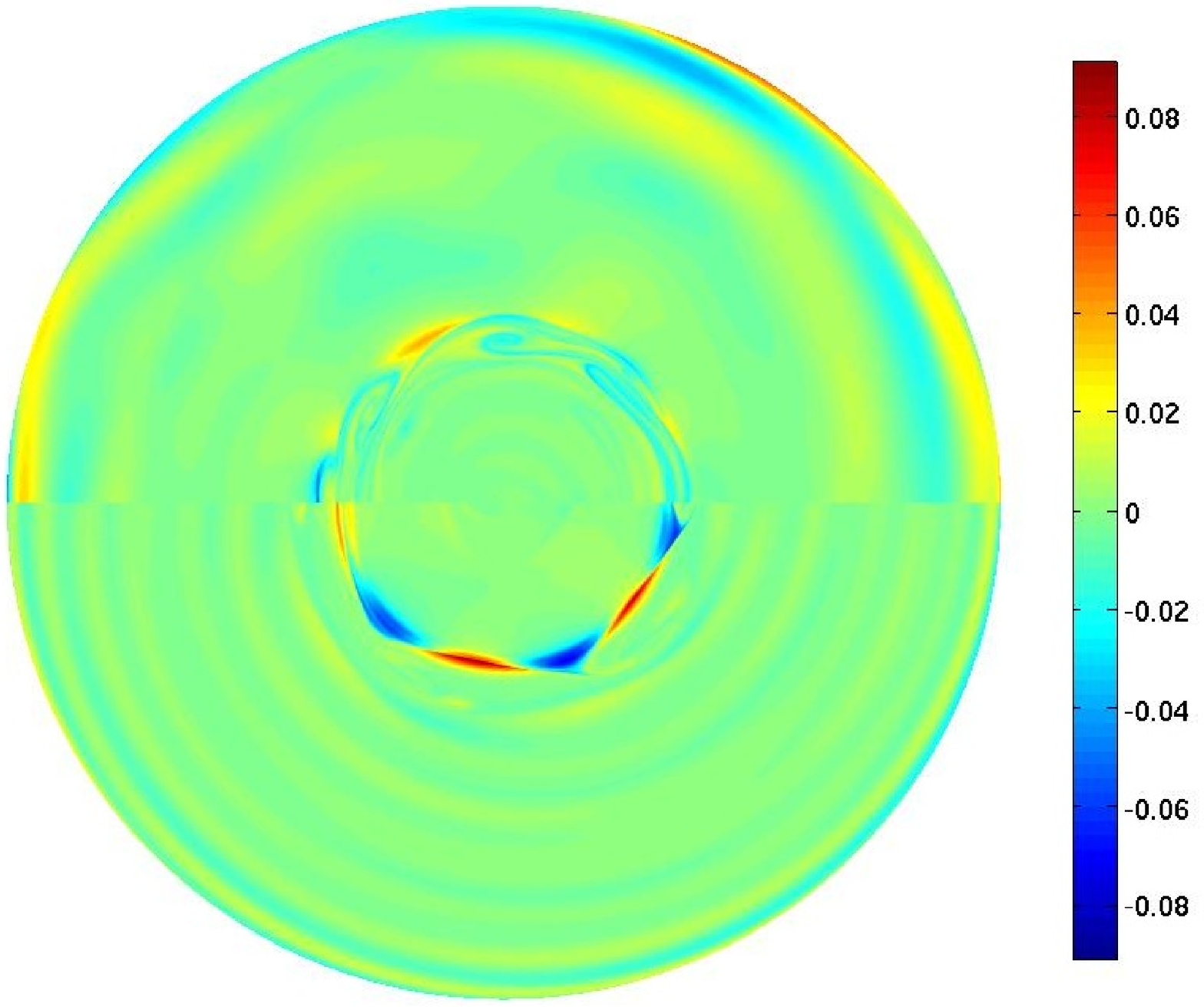} \\
\begin{picture}(0,0)
	\put(-165,160){a}
	\put(-165,25){b}
	\put(135,160){c}
	\put(135,25){d}
\end{picture}
\end{center}
\caption{$z$-vorticity maps in the equatorial plane. (a) and (b) : $E=10^{-6}$, $Ro = 0.0096$ and $Ro = -0.0111$ respectively. It shows the flow at the onset of hydrodynamic instabilities for both signs of the Rossby number. (c) and (d) : $E=10^{-8}$, $Ro=0.02$ and $Ro=-0.02$ respectively. It shows a typical view of the "turbulent" regime for Rossby numbers about 30 times critical. The color bar gives the local vorticity scale for (c) and (d) only.}
	\label{fig:vitesse}
\end{figure}

For low Rossby numbers, the split at the spherical boundary produces an internal shear layer in the fluid on a cylinder of radius $R_s$ aligned with the rotation axis. This geostrophic viscous layer consists of two nested layers of different widths as revealed by the asymptotic study of Stewartson \cite{ste66} and illustrated later by a numerical study of Dormy \emph{et al.} \cite{dor98}; a larger one of size $E^{1/4}$ which accommodates the jump in azimuthal velocity geostrophically and a narrower one of size $E^{1/3}$, ageostrophic, which corresponds to an axial jet insuring the mass conservation.

\begin{figure}
\includegraphics[width=11cm]{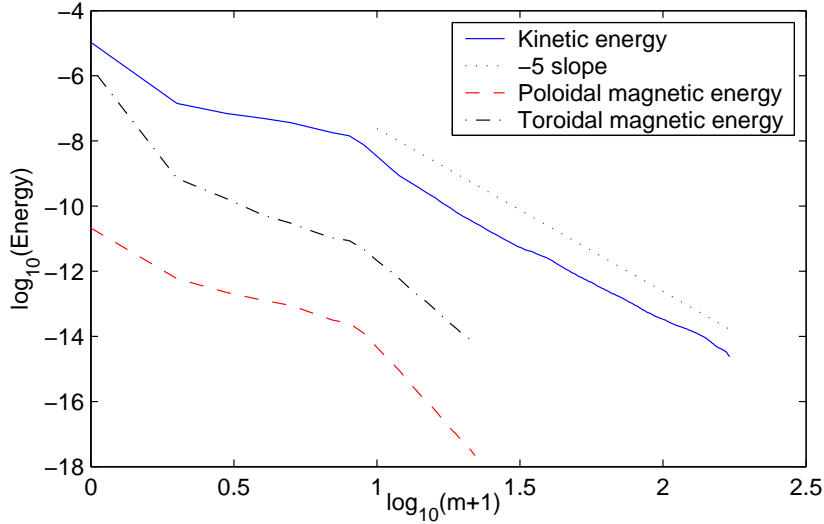}
	\caption{Spectra of the kinetic energy, and both toroidal and poloidal magnetic energy for $E=10^{-8}$, $Ro = 0.02$ (30 times critical) and $P_m = 5\,10^{-3}$ (equivalent to $Re=2\,10^6$).}
	\label{fig:spectra}
\end{figure}

In our previous study \cite{sch04}, we presented the QG model, which can reproduce only the $E^{1/4}$ layer, and we studied the linear perturbations of this geostrophic internal viscous layer. It becomes unstable when the Rossby number exceeds a critical value $Ro^c$ which varies as $\beta E^{1/2}$ \cite{sch04}. At the onset, the instability is a Rossby wave, azimuthal necklace of cyclones and anticyclones of size $E^{1/4}$ which propagates in the prograde direction as shown in figure \ref{fig:vitesse}ab.
Super rotation ($Ro>0$) generates a spiraling flow outside the shear layer while the flow is mainly located inside the shear layer for $Ro<0$.
For supercritical $Ro$, the flow exhibits larger vortices (fig. \ref{fig:vitesse}cd) which are time dependent but still drifting as Rossby waves. 
The flow stays mainly concentrated in the shear layer. 
Figure \ref{fig:spectra} shows the kinetic energy spectra $E(k)$ of this QG turbulent flow. It is very steep : $E(k) \sim k^{-5}$ which is the spectrum predicted by Rhines \cite{rhi75} for turbulence in presence of Rossby waves. This steep spectrum suggests that the small scales of the flow may be neglected in the induction equation.

\section{Dynamo action}

\begin{figure}
\includegraphics[width=14cm]{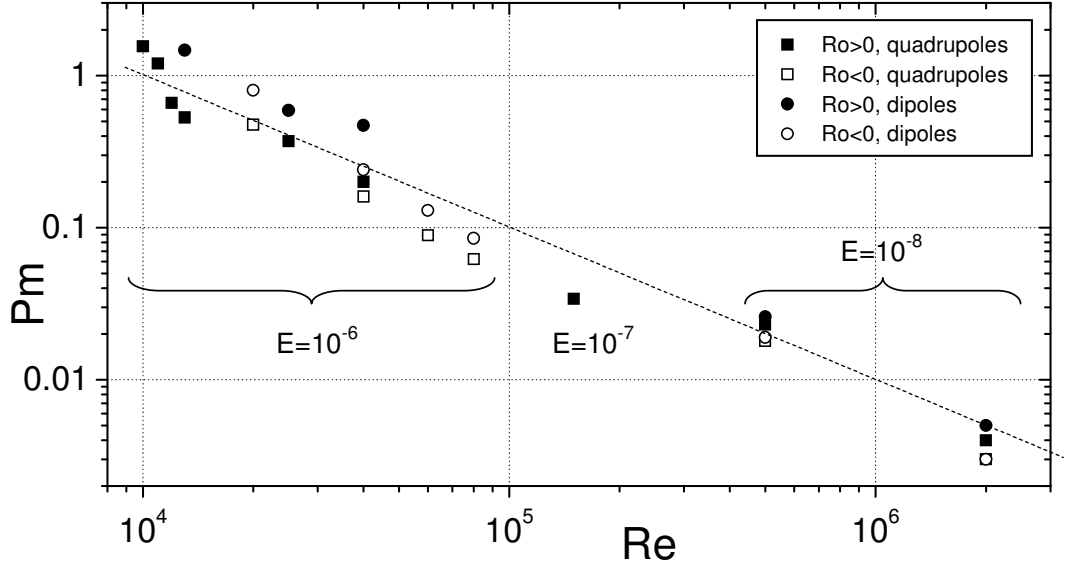}
	\caption{Dynamo onset for different parameters : Critical magnetic Prandtl number $P_m^c$ versus the absolute Reynolds number $Re = \left|Ro\right|E^{-1}$. Dipole and quadrupole thresholds are respectively denoted by circles and squares while solid and open symbols represent positive and negative differential rotation. All the points lie around the $R_m=10^{4}$ line.}
	\label{fig:dynRe}
\end{figure}

For a given Ekman number ($E=10^{-6}$ to $10^{-8}$), we vary the Rossby number $Ro$ from critical to a few times critical and we find the critical magnetic Prandtl number $P_m$ of the onset of dynamo action by trial and error tests.
As the flow is time dependent, we detect dynamo criticality on long term time variations of the magnetic energy. 
Unlike most of the kinematic dynamo models \cite{gub00}, a critical magnetic Prandtl number was found for every set of dimensionless numbers ($E,Ro$) we have computed.
In figure \ref{fig:dynRe}, we plot the calculated critical magnetic Prandtl number $P_m^c$ as function of the Reynolds number $Re=Ro/E$. As expected, we found that an increase of the forcing ($Ro$) for a given $E$ reduces the critical magnetic Prandtl number. A decrease of the critical magnetic Prandtl number is also observed as we lower the Ekman number. These two effects may be summarised by the use of the magnetic Reynolds number $R_m$. The data points in figure \ref{fig:dynRe} are roughly compatible with the line $R_m = 10^4$. A critical magnetic Reynolds number $R_m^c$ of $10^4$ is only indicative because the details of the flow generate large deviations (factor 3) from this simple law.
The minimum critical magnetic Prandtl number of $0.003$ has been found for $E= 10^{-8}$ and $Ro = 0.02$.

The critical magnetic Prandtl number is not independent of the sign of the differential rotation (sign of $Ro$). This is expected because the flow is quite different in the two cases as shown in figure \ref{fig:vitesse}. A negative differential rotation seems to lead to slightly lower dynamo thresholds.

\begin{figure}
\includegraphics[width=0.65\textwidth]{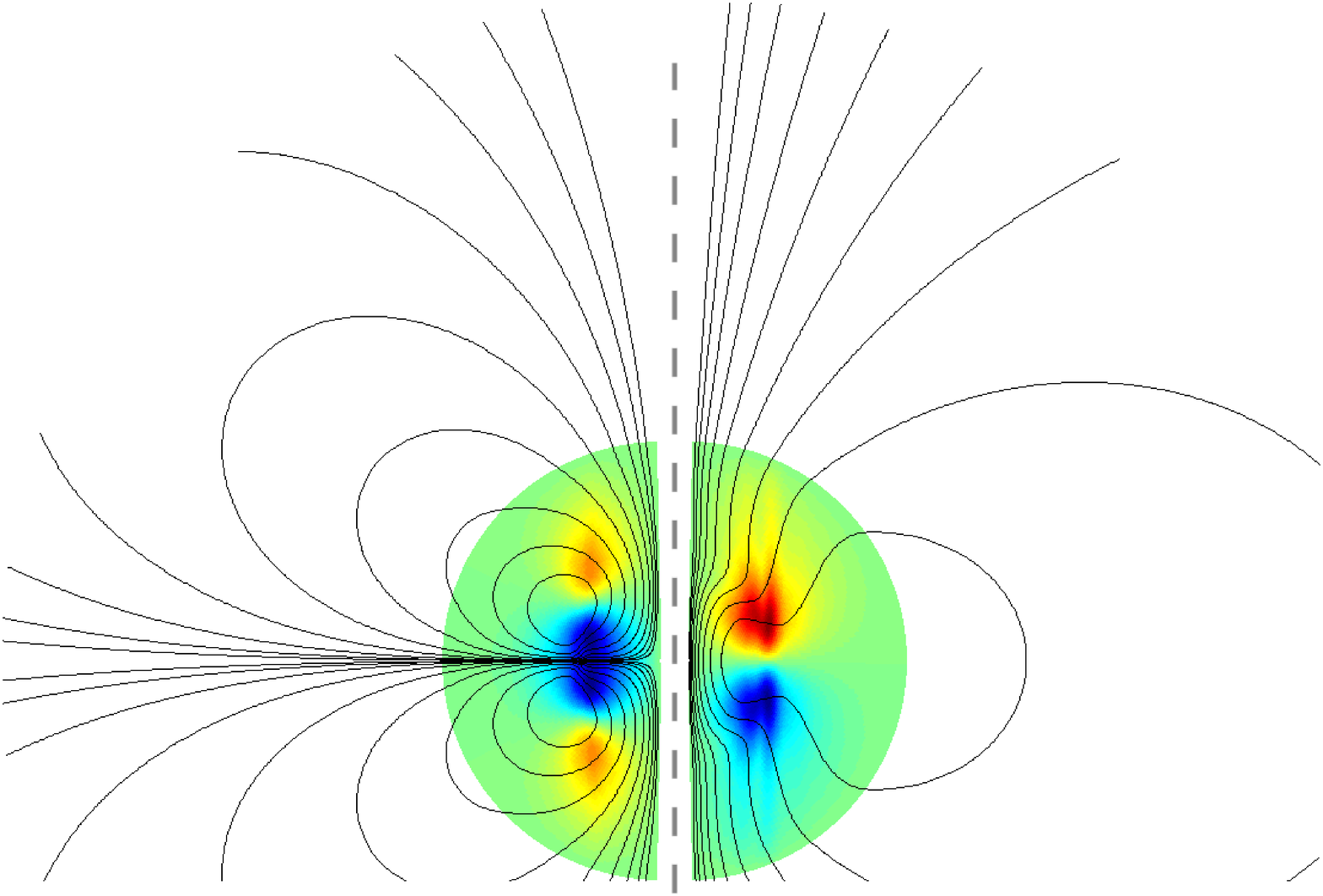}  \hspace{0.01\textwidth}
\includegraphics[width=0.3\textwidth]{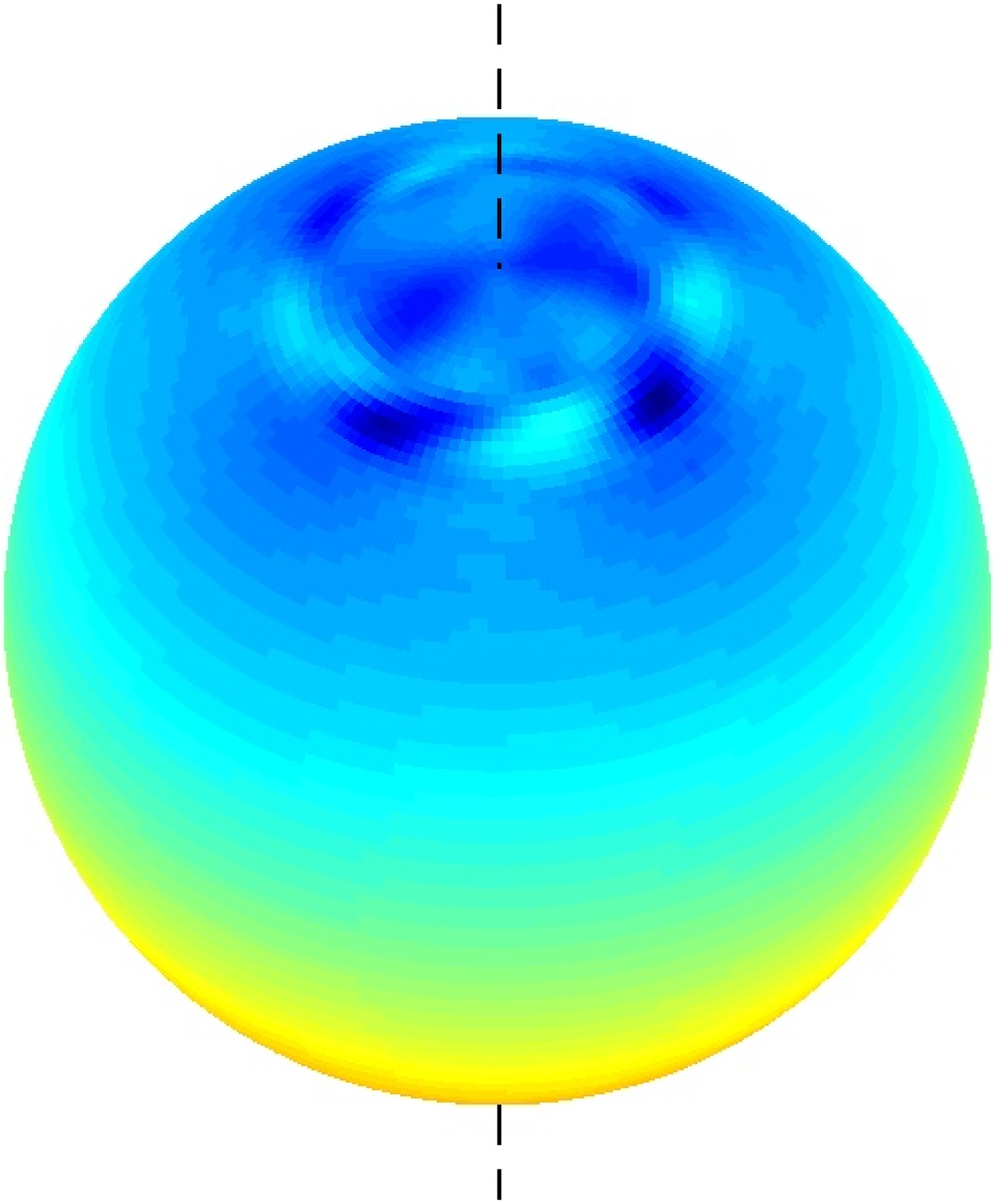} \\
\begin{picture}(0,0)
	\put(118,170){a}
	\put(138,170){b}
	\put(350,145){c}
\end{picture}
	\caption{Growing magnetic field in kinematic dynamos for $E = 10^{-8}$. (a) and (b) are meridian cut of the sphere showing the axisymmetric part of the magnetic field. The solid lines are the poloidal field lines and the color map represent the azimuthal field. (a) shows a quadrupole field obtained at $Ro=0.02$ and $P_m=0.005$. (b) shows a dipole field obtained at $Ro=-0.02$ and $P_m=0.003$. (c) is a spherical map of the radial magnetic field at the surface of the core, corresponding to case (b), the dashed line being the rotation axis. The corresponding vorticity fields are given in figure \ref{fig:vitesse}cd.}
	\label{fig:B}
\end{figure}

Antisymmetric axial velocities ($u_z(z) = -u_z(-z)$) and symmetric orthoaxial velocities ($u_{s,\phi}(z) = u_{s,\phi}(-z)$) generate two independent families of growing magnetic field in kinematic dynamos known as the dipole and quadrupole families \cite{rob72}.
The geometry of the two families are shown in figure \ref{fig:B}a and \ref{fig:B}b : the dipole family is dominated by an axial dipole, whereas the quadrupole family exhibit a strong axial quadrupole.
Each family has a different critical magnetic Prandtl number. As shown in figure 
\ref{fig:dynRe}, we found that the dipole family has always a larger critical magnetic Reynolds number than the quadrupole family. This result is quite different from the conclusion of the work of Sarson and Busse \cite{sar98}. Using Kumar and Roberts kinematic dynamos, they found that prograde spiraling of columns and prograde zonal flows favor dipole magnetic fields. 

In both families, the strongest magnetic fields are produced in the Stewartson shear layer deep inside the sphere. 
The typical spectra given in figure \ref{fig:spectra} show that the computed magnetic fields are dominated by both toroidal and axisymmetric components.
At the surface of the sphere (figure \ref{fig:B}c), the radial magnetic field is also mostly axisymmetric, and the non-axisymmetric part is clearly associated to the geostrophic vortices produced in the Stewartson shear layer.

The geometry of the magnetic field may be understood in term of $\alpha\Omega$ effects \cite{rob72,gub87}. A very large toroidal magnetic field compatible with the azimuthal flow is converted to a poloidal magnetic field by the columnar flow through an $\alpha$ effect. Any non azimuthal component of the magnetic field is transformed into an azimuthal component by the strong differential rotation in the Stewartson layer by $\Omega$ effect.

\begin{figure}
\begin{picture}(20,0)
	\put(130,-15){$\textrm{sgn}(Ro) \, \log_{10} (Ro/Ro^c)$}
	\put(0,100){$Dy$}
\end{picture}
\includegraphics[width=10cm]{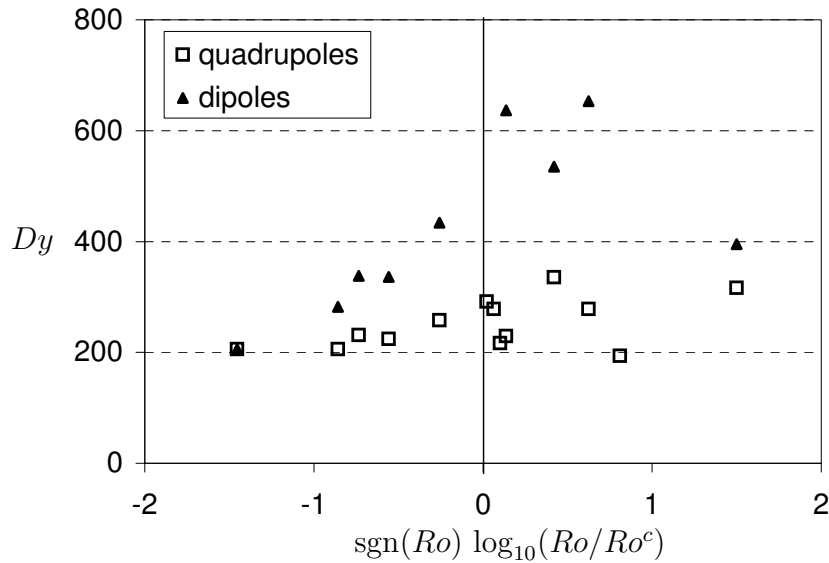}
\vspace{0.5cm}
	\caption{Onset of the dynamo instability for $E=10^{-6}$ to $10^{-8}$. The critical dynamo number $Dy^c$ is plotted versus $\log_{10}(Ro/Ro^c).\textrm{sgn}(Ro)$.}
	\label{fig:Dy}
\end{figure}

If we consider the magnetic Reynolds number associated to the shear flow $R_m^{\Omega} = Ro P_m E^{-1} $  and the magnetic Reynolds number based on the vertical velocity  $ R_m^{\alpha} = u_z P_m E^{-1} k^{-1}$ where $u_z$ is deduced from the calculation, as well as the Rhines wave number $k$ \cite{rhi75}. We may compute the dynamo number $Dy = \sqrt{R_m^{\Omega} R_m^{\alpha}}$ \cite{rob72}. Figure \ref{fig:Dy} shows that $Dy$ stays roughly constant (between 200 and 300) as the flow becomes more and more vigorous for the quadrupole family.
The dipole family seems more easy to excite for negative Rossby number. As proposed by Robert \cite{rob72}, this feature may indicate that we have $\alpha\Omega' <0$ (where $\Omega'$ is the radial derivative of $\Omega$) in the northern hemisphere for $Ro<0$.

Both the geometry and the onset in term of dynamo number $Dy$ indicate that we may look the Stewartson QG dynamo as an $\alpha\Omega$ dynamo where the $\Omega$ effect is produced by the azimuthal shear layer and the $\alpha$ effect by the vortex necklace.

Busse \cite{bus75} suggested that the Ekman pumping is important for the dynamo process~: although the $\beta$-effect produces axial velocities, they are out of phase with the axial vorticity at the onset of thermal convection in a rapidly rotating annulus and cannot contribute to the mean helicity, whereas axial velocities due to Ekman pumping are in phase with the axial vorticity.
However, the Ekman pumping flow is of order $E^{1/2}$, so that the dynamo process proposed by Busse becomes very weak when lowering the Ekman number.
In addition, when artificially removing the Ekman pumping flow in our dynamo simulations,
we still observe dynamo action with nearly the same threshold.
It seems to indicate that the $\beta$-effect alone may produce an efficient $\alpha$ effect, without requiring an Ekman pumping flow.

Furthermore, we have not been able to find a critical magnetic Prandtl number with a steady flow (either a time averaged flow or a flow with its time evolution stopped at a given time). 
It implies that the time evolution of the flow is a key ingredient for dynamo action in these quasi-geostrophic dynamos.
The propagation of the Rossby waves is required to put in phase the non axisymmetric magnetic fields and velocities in order to produce a axisymmetric poloidal magnetic field. This type of $\alpha$ effect was proposed in the model of Braginsky \cite{bra64}.
Currently, many dynamo experiments are designed with the help of numerical simulations (kinematic dynamos). Even if the flow is highly turbulent ($Re>10^6$), mean flow approaches are used for simplicity purposes to find the dynamo onset \cite{gai01,til02,dob03,mar03}. This method would fail in the case of Stewartson dynamos.

\begin{table}
	\label{tab:dyn_all}
	\begin{center}
		\begin{tabular}{ccccc}
		$E$ & $Ro$ & quadrupole $P_m^c$ & dipole $P_m^c$ \\
		\hline \\
$10^{-6} $ & $ 	1.00 \,10^{-2} $ & $ 	1.56  $ & \\
$10^{-6} $ & $ 	1.10\, 10^{-2} $ & $ 	1.2	  $ & \\
$10^{-6} $ & $ 	1.20\, 10^{-2} $ & $ 	0.66  $ & \\
$10^{-6} $ & $ 	1.30\, 10^{-2} $ & $ 	0.53  $ & $	1.47 $ \\
$10^{-6} $ & $ 	2.50\, 10^{-2} $ & $ 	0.37  $ & $ 0.59 $ \\
$10^{-6} $ & $ 	4.00\, 10^{-2} $ & $ 	0.2	  $ & $ 0.47 $  \\
$10^{-6} $ & $ 	-8.00\, 10^{-2} $ & $ 	0.062 $ & $	0.085 $ \\
$10^{-6} $ & $ 	-6.00\, 10^{-2} $ & $ 	0.089 $ & $	0.13 $ \\
$10^{-6} $ & $ 	-4.00\, 10^{-2} $ & $ 	0.16  $ & $	0.24 $ \\
$10^{-6} $ & $ 	-2.00\, 10^{-2} $ & $ 	0.475 $ & $	0.8 $ \\
$10^{-7} $ & $ 	1.50\, 10^{-2} $ & $ 	0.03  $ & \\
$10^{-8} $ & $ 	2.00\, 10^{-2} $ & $ 	0.004 $ & $	0.005 $  \\
$10^{-8} $ & $ 	-2.00\, 10^{-2} $ & $ 	0.003 $ & $	0003 $ \\
		\end{tabular}
        \caption{Table of the critical magnetic Prandtl numbers for the different calculations.}
	\end{center}
\end{table}

\subsection{Oscillating solution}

\begin{figure}
        \includegraphics[width=16cm]{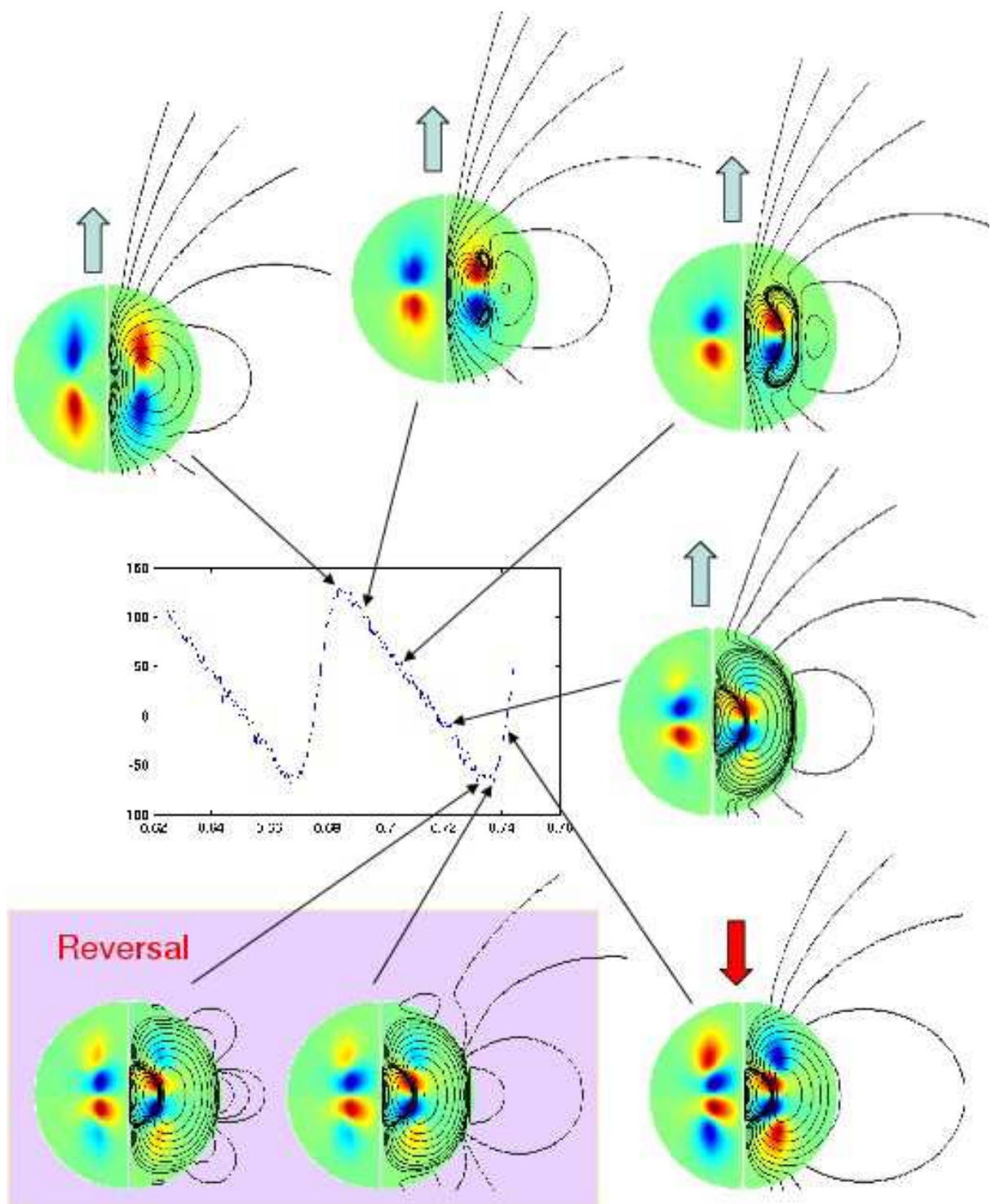}        
        \caption{Magnetic field reversal observed at $E=10^{-6}$, $Ro=-0.08$ and $P_m=0.1$. The graph shows the evolution of the growth rate of the magnetic energy as a function of time (in magnetic diffusion time units).}
        \label{fig:dynRev}
\end{figure}

As in many $\alpha\Omega$ dynamo \cite{rob72}, we sometimes obtain a time oscillating solution for the Stewartson dynamo. Dipole solutions for $E=10^{-6}$ do exhibit such a behavior.
The growth rate of one of these dynamos is plotted on figure \ref{fig:dynRev}, showing three time scales : the smallest one is the time scale of the velocity fluctuations. The intermediate time scale is the time needed for the growth rate to go from its minimum value to its maximum. The large time scale is the period of oscillation, not linked to any time scale of the flow.

In the context of kinematic dynamos, this behavior correspond to a complex eigenvalue of the linear set of equation \cite{rob72,gub87,dud89}. We use the same formalism to explain our result. It may be explained by two coupled magnetic modes $B_1$ and $B_2$. Let assume that the induction equation may be approximated by the following system :
\begin{eqnarray}
\label{eq:2modes}
	\frac{d B_1}{d t} & = & \lambda_1 \, B_1 + K_{12} \, B_2 \\
	\frac{d B_2}{d t} & = & \lambda_2 \, B_2 - K_{21} \, B_1
\end{eqnarray}
with all real values.
For low coupling ($K_{12}K_{21} < (\lambda_1 -\lambda_2)^2/2$) the eigenvalues of this system are real, so that the growing solution is a combination of $B_1$ and $B_2$. This is the case for the quadrupole family at $Ro<0$.
However, when the coupling $K_{12}K_{21}$ is sufficiently strong, the eigenvalues are complex and as a result the growing magnetic field oscillates periodically between $B_1$ and $B_2$.
The intermediate time scale (time for the growth rate to go from its minimum to its maximum) is very close to the phase shift between $B_1$ and $B_2$, and one of the two modes is dominant near the minimum of the growth rate cycle, while the other one is dominant near the maximum, with growth rate close to $\lambda_1$ and $\lambda_2$.

The reversal process at work in our simulations is a smooth periodic evolution of the magnetic field, but at the surface it appears to be a sudden sign reversal.
In fact, a reversed poloidal magnetic field is slowly growing inside the Stewartson layer, moving away the initial poloidal magnetic field until it reaches the outer boundary. Then, the reversed dipole magnetic field suddenly appears at the surface and ultimately the poloidal field reverses at the center. During the time oscillation, the axisymmetric toroidal magnetic field patches in the Stewartson layer migrate toward the equator as reversed polarity toroidal fields are formed at higher latitudes. This migration could be understood in terms of Parker dynamo waves \cite{par55,rob72}.

\section{Conclusion}

In summary, we have computed a quasi-geostrophic dynamo based on a Stewartson shear layer flow. The scale separation approach works because the small scales of the flow in our rotating sphere are negligible (very steep kinetic energy spectrum $E(k) \sim k^{-5}$).
Our preliminary results may be interpreted in terms of $\alpha\Omega$ dynamo. The $\Omega$ effect is done by the shear of the Stewartson layer itself whereas the $\alpha$ effect is produced by vortices associated with the Rossby waves due to the instability of the shear layer. These understandings are very encouraging for our on-going experimental modeling of the geodynamo. As described in Cardin et al. \cite{car02}, we are building a spherical Couette experiment using liquid sodium which may validate and enlarge our present numerical findings.

For the first time, we have computed a spherical dynamo with a very low magnetic Prandtl number ($<10^{-2}$) and a very low Ekman number ($10^{-8}$) (corresponding to a very high Reynolds number $Re>10^{6}$). Even if our dimensionless parameters stay far away from parameters of planetary cores, our calculations use dimensionless numbers which are in the correct asymptotic regime for the modeling of the geodynamo. The key ingredients of our approach is to take into account a specific property of the rotating fluid (QG) which allows us to use a 2D model to compute the flow evolution, and the separation of scales between the magnetic field and the velocity field, allowing us to use a coarse 3D mesh for the magnetic field.

We also showed that in the case studied in this paper, the mean flow or the static flow fails to produce a dynamo while the fully resolved time-dependent flow succeeds.
Indeed, the time evolution of the flow and the $\beta$ effect are key ingredients for dynamo action in our models, while the Ekman pumping can be neglected without losing the dynamo effect.

The next step will be to add the Lorentz force in the QG equation to compute saturated dynamos. One of the difficulty is to compute the action of the large magnetic field on the small scale motions of the fluid. Preliminary results are encouraging and exhibit saturated dynamos very close to the kinematic dynamos described here. 

A quasi geostrophic approach could also be used to build thermal convective dynamos. A zonal geostrophic flow is produced by the Reynolds stress of the thermal columns \cite{aub01,chr02} but its amplitude is much lower compared to the differential rotation imposed in the Stewartson problem. Would it be enough to start an Stewartson dynamo type? for what forcing? Would it work for very low Ekman and magnetic Prandtl numbers?

\vspace{1cm}
{\it Aknowledgements}: Calculations were performed at SCCI (Observatoire de Grenoble) and at IDRIS (CNRS). This work has been supported by the programme "DyETI" of CNRS/INSU. We wish to thank Dominique Jault and Henri-Claude Nataf
for very useful comments.

\bibliographystyle{unsrt}
\bibliography{article_vhal}

\end{document}